\documentclass[useAMS,usenatbib]{mn2e}
\usepackage{graphics,epsfig,aas_macros}
\usepackage[normalem]{ulem}
\usepackage[dvipsnames]{color}
\usepackage[]{inputenc,amssymb}
\usepackage{amsmath}
\usepackage{color}
\usepackage{textcomp}
\usepackage{array}
\usepackage{textcase}
\usepackage{titlesec}
\usepackage{etoolbox}

\makeatletter
\patchcmd{\ttlh@hang}{\parindent\z@}{\parindent\z@\leavevmode}{}{}
\patchcmd{\ttlh@hang}{\noindent}{}{}{}
\makeatother
\renewcommand\footnoterule{\kern 5pt \hrule width 2in \kern 5.0pt}

\titleformat{\paragraph}
{\it\normalsize\normalsize}{\theparagraph}{1em}{}
\titlespacing*{\paragraph}
{0pt}{3.25ex plus 1ex minus .2ex}{1.5ex plus .2ex}

\newcolumntype{P}[1]{>{\centering\arraybackslash}p{#1}}

\usepackage[toc,page]{appendix}

\def \be{\begin{equation}}
\def \ee{\end{equation}}
\def \bea{\begin{eqnarray}}
\def \eea{\end{eqnarray}}

\title[X-ray AGN]{{Constraining the X-ray AGN halo occupation distribution: implications for eROSITA}}

\author [Singh, Refregier, Majumdar \& Nath]
{Priyanka Singh$^{1}$ \thanks{priyankas@rri.res.in}, 
 Alexandre Refregier$^2$, 
Subhabrata Majumdar$^3$ 
 and Biman B. Nath$^1$ 
 \\
$^1$ Raman Research Institute, Bangalore 560080, India \\
$^2$ Institute for Astronomy, Department of Physics, ETH Z\"{u}rich, Wolfgang-Pauli-Strasse 27, CH-8093 Z\"{u}rich, Switzerland\\
$^3$ Tata Institute of Fundamental Research, Mumbai 400005, India
}

\begin{document}


\maketitle

\label{firstpage}

\begin{abstract}
The X-ray emission from active galactic nucleus (AGN) is a major component of extragalactic X-ray sky.
In this paper, we use the X-ray luminosity function (XLF) and halo occupation distribution (HOD)
formalism to construct a halo model for the X-ray emission from AGNs. 
Verifying that the two inputs (XLF and HOD) are in agreement with each other, we compute the auto-correlation power
spectrum in the soft X-ray band (0.5-2 keV) due to the AGNs potentially resolved by eROSITA
(extended ROentgen Survey with an Imaging Telescope Array) mission and explore the 
redshift and mass dependence of the power spectrum.
Studying the relative contribution of the Poisson and the clustering terms to the total power,
we find that at multipoles $l\lesssim 1000$ (i.e. large scales), the 
clustering term is larger than the Poisson term. 
We also forecast the potential of X-ray auto-correlation power spectrum and 
X-ray-lensing cross-correlation power spectrum using eROSITA and eROSITA-LSST (Large Synoptic Survey Telescope)
surveys, respectively, to constrain the HOD parameters and their redshift evolution.
In addition, we compute the power spectrum of the AGNs lying below 
the flux resolution limit of eROSITA, which is essential to understand in order to extract the X-ray signal
from the hot diffuse gas present in galaxies and clusters.
\end{abstract}


\section{Introduction}
\label{sec-intro}
Almost every galaxy with a central bulge contains a supermassive black hole (SMBH) 
($M_{\rm SMBH} \gtrsim 10^6 M_{\odot}$) at its centre \citep{kormendy95, ferrarese2000}.
The SMBH acquires a large mass through accretion of matter from its surrounding and the merger
of the host galaxies. These processes trigger the active galactic nucleus (AGN) phase of the SMBH during which 
the accretion rates are high and the accreting material emits a large fraction of its rest energy.
The luminosity of the AGN may surpass the total light from the galaxy and drive strong galactic outflows.
The evidence of such a feedback from the AGN is present in massive galaxies, groups and clusters. 
Simulations \citep{omma04, springel05, mcnamara07, puchwein08, battaglia10, maccarthy10}
as well as analytical studies \citep{valageas99,
bower01, cavaliere02, mahavir13}
show that, in the absence of any AGN feedback, it is difficult to reproduce various observations such as the
gas mass fractions, X-ray luminosity scaling relations, heating of cooling flows of galaxy clusters and
gas pressure profiles. The cluster gas entropy also holds key to AGN feedback, which can be directly 
linked to the non-gravitational energy deposited and remaining in the intracluster
medium (ICM; \citealt{chaudhuri12, chaudhuri13, iqbal16}).
These studies give indirect but strong evidence of the presence of AGN and support the importance 
of AGN driven feedback in the evolution of galaxies.

In addition to giving rise to the outflows, AGNs are strong X-rays emitters and form a
dominant part of the extragalactic X-ray sky. 
eROSITA (extended ROentgen Survey with an Imaging Telescope Array) which is a future X-ray satellite with all 
sky coverage, will provide a large sample of X-ray AGNs \citep{merloni12, kolodzig13b}. It will also cover a significant 
range of angular scales due to its large sky coverage and improved angular resolution ($\sim 30''$ 
in 0.5-2 keV band)
compared to ROSAT all-sky survey, which was the last all-sky X-ray survey with 
mean point spread function $\sim 2'$ \citep{labarbera09}.

In this paper, we first compute the angular auto-correlation power spectrum of the X-ray emission coming from AGNs, 
that are expected to be resolved by eROSITA survey,
using a halo model approach. There are two main ingredients of this approach.
(1) A model that describes the X-ray emission from AGNs. It can be obtained using the 
X-ray luminosity function (XLF) of AGNs which represents the luminosity distribution
of AGNs as a function of redshift. Here we use the XLF given in \citealt{aird15} (hereafter A15).
(2) We also need a model that describes how AGNs populate dark matter haloes (DMH).
Large efforts have been invested into probing how the DMH affect the distribution
and energetics of AGNs \citep{croom04, gilli05, krumpe12, white12, koutoulidis13, mountrichas13, allevato14, gatti16}.
The clustering measurements of X-ray AGNs suggest that AGNs 
occupy haloes in the mass range $\sim 10^{12.5}-10^{13.5} h^{-1} M_{\odot}$ \citep{coil09, cappelluti10, allevato11, leau15}.
AGN clustering measurements can 
also be used to construct the halo occupation distribution (HOD) model of AGNs
(\citealt{miyaji11, allevato12, richardson13}, hereafter R13). In this formalism,
the number of AGN (central+satellite) is modelled in terms of the mass of the host DMH. The 
HOD analysis describes how AGNs populate the DMH, which can be used to construct the AGN power
spectrum. Specifically, we use the HOD model described in R13.
This HOD model, however, lacks a redshift dependence, which 
is expected from the redshift dependence of AGN XLF and the host halo properties.
Here, we study the constraints that can be put on the redshift evolution of the HOD model with eROSITA and 
Large Synoptic Survey Telescope (LSST, \citealt{lsst09}) 
using the X-ray auto-correlation power spectrum and X-ray-lensing cross-correlation power spectrum of the resolved AGN.
The choice of the X-ray-lensing cross-correlation power spectrum is motivated by the fact that the 
HOD formalism connects the distribution of the AGNs with the host DMH mass and lensing signal
is direct tracer of the mass.

Other important components of the X-ray sky are X-ray binaries, the hot gas present in our own Galaxy and the 
extragalactic hot gas. 
Amongst these X-ray sources, the large hot gas reservoir ($T_{\rm vir} \sim 10^7 K$) filling the space between
the galaxies in the clusters, known as the ICM, has been observed in its X-ray 
emission for a long time \citep{reichert81, jones84, raymont85, oukbir97, diego03, diego03b, cavagnolo09, hurier15}. 
However, the hot gas ($T_{\rm vir} > 10^6 K$) present in the form of circumgalactic medium (CGM)
in massive galaxies ($M_h \sim 10^{12} - 10^{13} h^{-1} M_{\odot}$; \citealt{birn03, keres05, singh15a}) 
is less explored in X-rays
due to its fainter X-ray emission. Some of the recent observations
and studies \citep{putman09, anderson11, dai12, anderson13, bogdan13a, bogdan13b, putman12, gatto13} 
indicate that the CGM can account for a good fraction of the baryons in these galaxies.
The X-ray emission from the CGM, therefore, is a promising tool to put strong
constraints on the distribution and energetics of the gas \citep{singh15b} with eROSITA.
At energies above 2 keV, the extragalactic point sources like AGN completely dominate the X-ray sky
\citep{lehmann01, kim07}.
Even below 2 keV, where the X-ray emission from the hot gas in the ICM and CGM
is significant, the major contribution 
 to the observed X-ray sky comes from AGNs \citep{soltan07}.
Therefore, studying the X-ray emission from the AGN is crucial to understand the origin and evolution of the AGN as well
 as to extract the X-ray signal from the subdominant components.
We thus also compute the angular power spectrum of the unresolved AGNs which are expected to contribute to the
diffuse X-ray background of eROSITA and contaminate the angular power spectrum due the ICM/CGM in the 0.5-2 keV X-ray band.

This paper is organised as follows. In section-\ref{sec-resolved}, we 
describe the methodology and various ingredients required for 
 the calculation of X-ray power spectrum from the resolved AGNs, and 
show the redshift and halo mass dependence of the power spectrum. In section-\ref{sec-forecast}, we forecast the 
constraints on the HOD parameters and their redshift evolution. 
In section-\ref{sec-kxcross}, we describe the X-ray-lensing cross-power spectrum and its power to 
put stringent constraints on HOD model parameters.
In section-\ref{sec-unresolved}, we compute
the X-ray auto-correlation power spectrum due to the unresolved AGNs, its redshift and halo mass dependence.
Finally, we summarise our main conclusions in section-\ref{sec-conc}. The cosmological parameters used in this 
paper are taken from \cite{planck15}.

\begin{figure*}
\begin{center}
\includegraphics[width=8.5cm,angle=0.0 ]{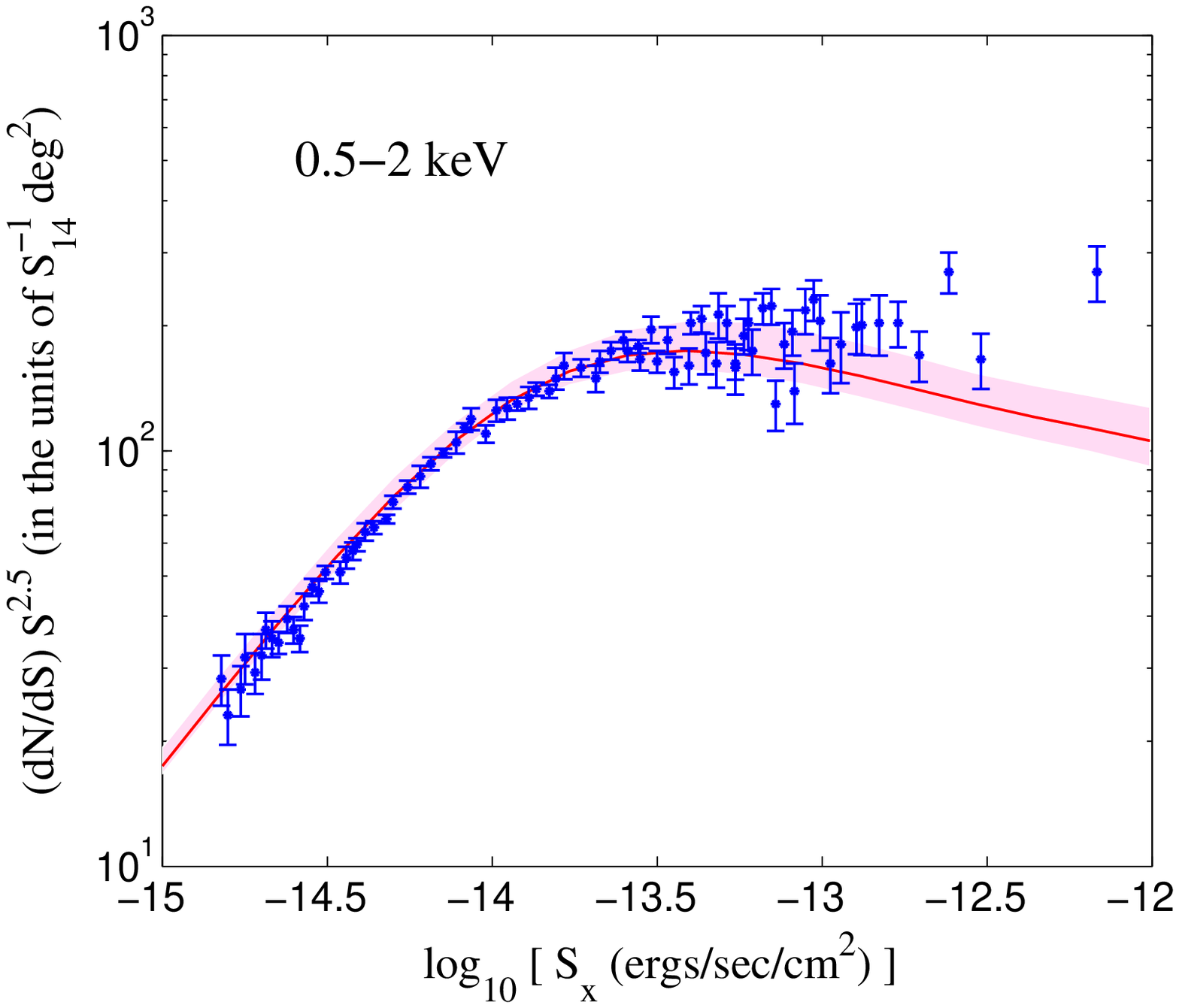}
\includegraphics[width=8.5cm,angle=0.0 ]{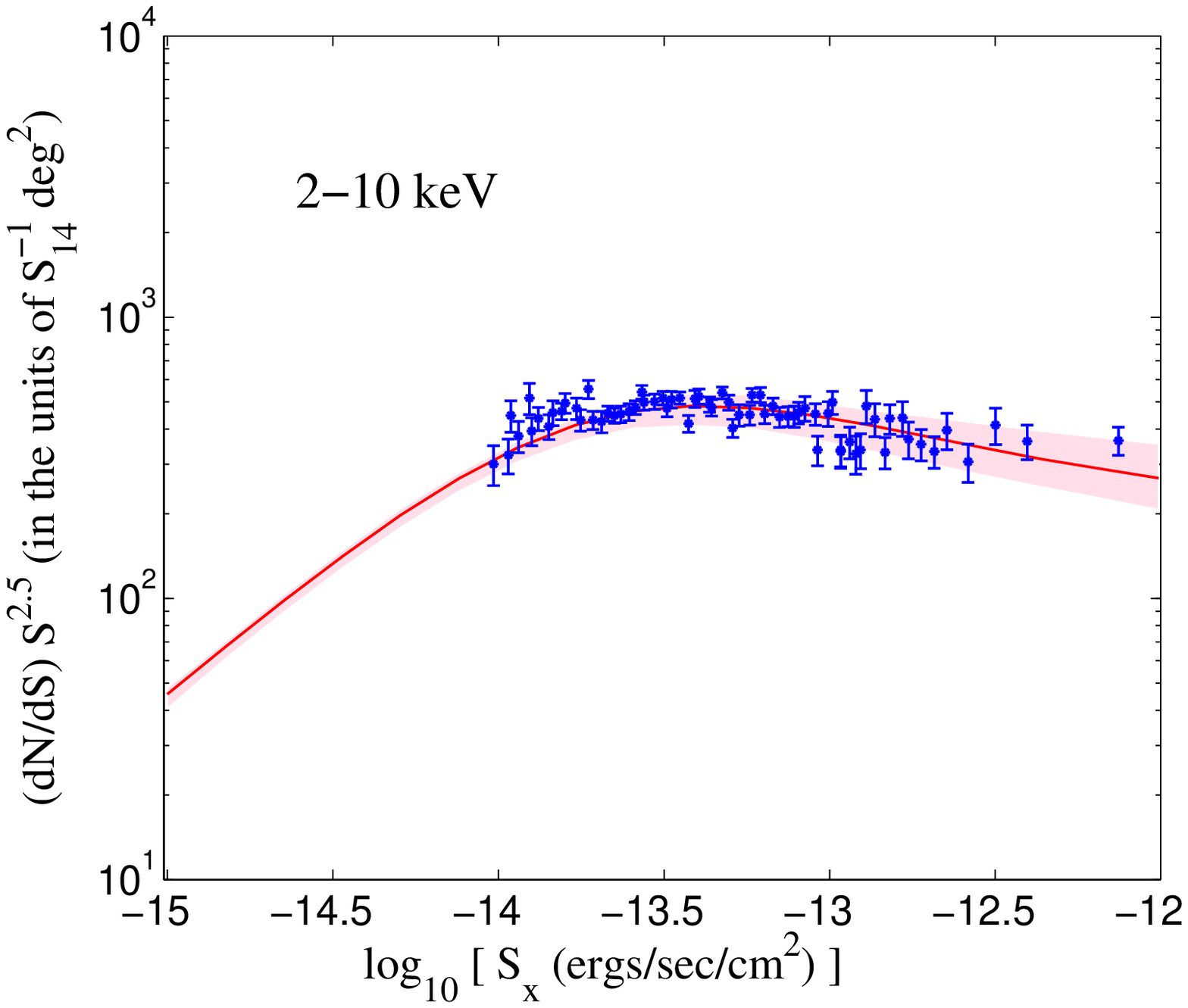}
\end{center}
\caption{The observed differential number counts of X-ray AGNs (blue points with error bars) from \citet{mateos08}
 against the prediction of the XLF (solid red line) in the soft (left-hand panel) and the hard (right-hand panel) X-ray bands.
 In both the panels, the shaded region represents the 3-$\sigma$ uncertainty in the prediction of the XLF.}
\label{fig-dnds}
\end{figure*}

\section{Auto-correlation power spectrum of the resolved AGNs}
\label{sec-resolved}
Given a signal $x(\theta,\phi)$, its angular power spectrum is given by (eg. \citealt{hivon02, molinari14}),
\begin{equation}
C_l=\frac{1}{2l+1} \sum_{m=-l} ^{m=l} |a_{lm}|^2 
\label{eqn-estim}
\end{equation}
where, $a_{lm}=\int d \Omega\, x(\theta,\phi)\, w(\theta,\phi)\, Y^{*} _{lm}(\theta,\phi)$, $w(\theta,\phi)$ is the mask 
(decided by the sky coverage of the survey with low foreground and low instrumental noise) and $Y^{*} _{lm}$ are the spherical
harmonic functions. There are also other corrections applied to the above power spectrum due to the partial sky coverage
of the survey, which is generally the case.

Presently, due to the small sky coverage of the ongoing X-ray surveys, 
the clustering of AGNs are usually studied only through their two-point correlation function (2PCF).
But, with the launch of eROSITA, it will be possible to explore the power spectrum of these
AGNs in great detail. Therefore, we choose to deal with the number-weighted power spectrum of AGNs that are
expected to be resolved by eROSITA i.e. AGNs lying above the flux resolution limit of eROSITA.
We assume that all AGNs above the flux limits $\sim 1.5\times10^{-14}$ and 
$1.8\times10^{-13} \rm erg\,\, s^{-1} cm^{-2}$ in 0.5-2 keV (soft) and 2-10 keV (hard) X-ray bands, respectively,
will be resolved by eROSITA.

For resolved AGNs, $x(\theta,\phi)$ represents the AGN number counts, whereas in the case of unresolved AGNs, it represents 
unresolved X-ray flux. In reality, the observed signal also contains contribution from many other sources and 
one has to carefully remove the noise to get the desired power spectrum. In this paper, we show the analytical 
estimate of the angular power spectrum calculated using halo model approach. In this approach,
the AGN power spectrum is represented by the total contribution of the AGNs 
residing in a halo, convolved with the DMH mass function, integrated over mass and redshift, as a function
of mutipole $l$.
The AGN angular auto-correlation power spectrum is the sum of two terms,
\begin{equation}
  C^{\rm AGN}_l=C^{\rm AGN,P}_{l}+C^{\rm AGN,C}_{l}
\end{equation}
where $C^{\rm AGN,P}_{l}$ and $C^{\rm AGN,C}_{l}$ are the Poisson and clustering terms, respectively.\\

\subsection{Poisson term}
The Poisson term (independent of $l$; also known as shot noise) of the AGN angular auto-correlation power spectrum is
given by
\begin{equation}
 C^{\rm AGN}_{P}=\int dz \frac{dV}{dzd\Omega} \int d\log L_X \phi_{\rm AGN}(L_X,z) 
 \label{eqn-clp}
\end{equation}
The above equation also represents the number of AGNs per unit solid angle.
Here $\phi_{\rm AGN}(L_X,z)$ is the XLF and $L_X$ is the X-ray luminosity of the AGN (in the hard X-ray band)
related to the observed X-ray flux (in the soft X-ray band) $S_X$ by,

\begin{equation}
 L_X=\frac{4 \pi d^2_L(z) S_X}{(1+z)^{2-\Gamma}} 
 \frac{E^{2-\Gamma}_{\rm max,\, RF}-E^{2-\Gamma}_{\rm min,\, RF}}{E^{2-\Gamma}_{\rm max,\, obs}-E^{2-\Gamma}_{\rm min,\, obs}}
 \label{eqn-bandcorrec}
\end{equation}

where $d_L(z)$ is the luminosity distance, $\Gamma$ is the AGN spectral index 
(assuming that the AGN X-ray emission follows a power law spectrum), $E_{\rm max,\, RF}$ and $E_{\rm min,\, RF}$
are the upper and lower limit of the X-ray band in the rest frame of the AGN, respectively, whereas, 
$E_{\rm max,\, obs}$ and $E_{\rm min,\, obs}$ correspond to the observed X-ray band.
The lower luminosity limit in above integral is determined by the sensitivity limit of the telescopes 
, which is $\sim 1.5\times10^{-14} \rm erg\,\, s^{-1} cm^{-2}$ and 
$1.8\times10^{-13} \rm erg\,\, s^{-1} cm^{-2}$ for soft and hard X-ray bands, respectively.

\subsection{AGN XLF}
\label{sec-xlf}
The AGN XLF is defined as the comoving number density of AGNs per unit logarithmic
X-ray luminosity, i.e. $\phi_{\rm AGN}(L_X,z)=dn/d\log_{10} L_X$. 
To compute it, we use the Luminosity And Density Evolution (LADE) model (see A15 for the details of the model),
which gives
\begin{equation}
 \phi_{\rm AGN}(L_X,z)=K(z) \Bigl[ \Bigl(\frac{L_X}{L_*(z)}\Bigr)^{\gamma_1} + 
 \Bigl(\frac{L_X}{L_*(z)}\Bigr)^{\gamma_2} \Bigr ]^{-1},
\end{equation}
where
$ K(z)=K_0 \times 10^{d(1+z)}$
and $L_*(z)=L_0 \Bigl[ \Bigl(\frac{1+z_c}{1+z}\Bigr)^{p_1}+\Bigl(\frac{1+z_c}{1+z}\Bigr)^{p_2} \Bigr ]^{-1}$.
The values of the model parameters are shown in table-\ref{tab-lade}. Note that this XLF corresponds
to luminosities integrated in the hard X-ray band in the rest frame of the AGN.
Therefore, one has to take into account the band correction (see equation-\ref{eqn-bandcorrec})
to estimate the XLF in different X-ray bands.

The differential number counts i.e. the number of AGNs per unit flux and solid angle in the soft band is
\begin{equation}
 \frac{dN}{dS_X d\Omega}=\frac{1}{S_X \ln10} \int dz \frac{dV}{dzd\Omega} \phi_{\rm AGN}(L_X,z),
\end{equation}
where $dV/dzd\Omega$ is the differential comoving volume.
We compare the AGN differential number counts calculated using the LADE model with the observed number counts \citep{mateos08}
for  soft as well as  hard X-ray band 
 in figure-\ref{fig-dnds}. 
We take $\Gamma =$1.7 for soft and 2.0 for the hard X-ray band.
There is a good agreement between the observed AGN counts and the LADE model.

\begin{table}
\caption{LADE model parameters.}
\centering 
\begin{tabular}{ c c c }
 \hline 
 Parameter & Soft X-ray band & Hard X-ray band \\[0.5ex]
 \hline \\
 
$\log K_0 (\rm Mpc^{-3})$ & -4.28 & -4.03\\
$\log L_0 (\rm erg\hspace{1 mm} \rm s^{-1})$ & 44.93 & 44.84 \\
$\gamma_1$ & 0.44 & 0.48\\
$\gamma_2$ & 2.18 & 2.27 \\
$p_1$ & 3.39 & 3.87 \\
$p_2$ & -3.58 & -2.12 \\
$z_c$ & 2.31 & 2.00 \\
$d$ & -0.22 & -0.19 \\
\hline
 \end{tabular}
\label{tab-lade}
\end{table}

\begin{figure}
\begin{center}
\includegraphics[width=8.2cm,angle=0.0 ]{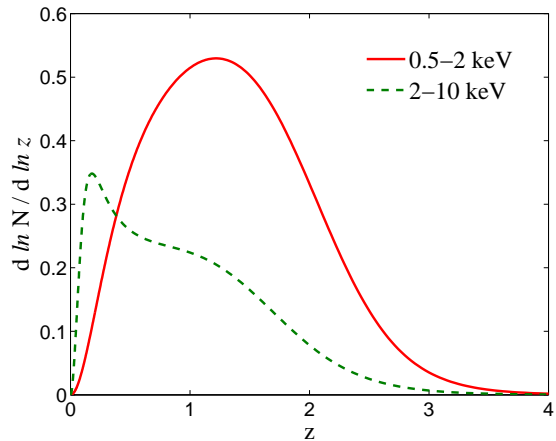}
\end{center}
\caption {The differential redshift distribution of the AGNs expected to be resolved by eROSITA all-sky survey
in the soft (solid red line)
and the hard (dashed green line) X-ray bands.}
\label{fig-dndz}
\end{figure}

In figure-\ref{fig-dndz}, we show the redshift distribution of AGNs that 
are expected to be observed
by eROSITA in the soft as well as the hard 
X-ray band. The soft and hard band AGN number counts get maximum contribution
from AGNs near $z\sim$ 1.2 and 0.2, respectively. The hard band AGNs are less concentrated near 
the peak than the soft band AGNs with a hint of another peak near z$\sim$ 1.2.
Beyond $z\sim$ 4, the number of resolved AGNs becomes negligible.

\begin{figure}
\begin{center}
\includegraphics[width=8.2cm,angle=0.0 ]{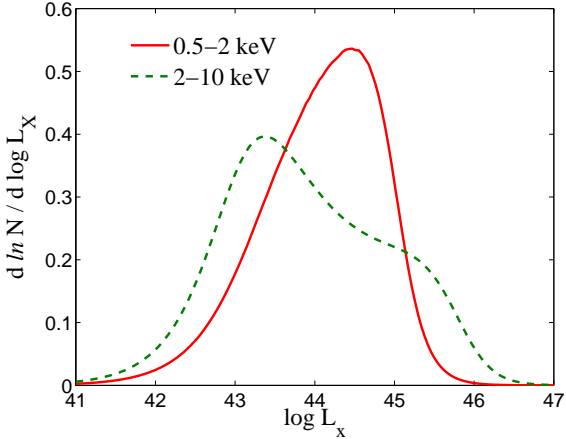}
\end{center}
\caption {Same as figure-\ref{fig-dndz} but for the luminosity distribution of the resolved AGNs.}
\label{fig-dndl}
\end{figure}

In figure-\ref{fig-dndl}, we show the luminosity distribution of the resolved AGNs in the soft and 
hard X-ray bands. In the soft band, the luminosity distribution peaks at $L_X \sim 10^{44.5}
\rm erg\, s^{-1}$. In the hard band, the peak is at smaller X-ray luminosity,
$L_X \sim 10^{43.4} \rm erg\, s^{-1}$. Again, the hard X-ray AGNs have a much broader luminosity
distribution compared to the soft X-ray AGNs. In both the bands, negligible fraction of the 
total resolved AGN population lies outside the luminosity range $L_X \sim 10^{41}$-
$10^{47}$ erg s$^{-1}$.

Note that, the prediction of redshift and luminosity dependence of the AGNs crucially depend on the choice
of the XLF. Different XLFs, though broadly consistent with each other, give rise to different shapes and peak 
values of the AGN redshift and luminosity distributions (see figure-10 of \citealt{kolodzig13a}). Also, our default 
XLF (A15) describes the AGN in 2-10 keV rest frame and 
for simplicity
we use the power law approximation to get the soft band XLF, neglecting the effect of evolving fraction
of absorbed and unabsorbed AGNs, which may affect the overall redshift distribution of the AGNs.
Our choice of XLF is motivated by the reasonable reproduction of AGN number counts (shown in figure-\ref{fig-dnds})
and its consistency with the HOD model used in this paper (discussed later in section-\ref{sec-2hc}).
We have also found a good agreement between the predictions of redshift distribution of the soft band AGNs using A15 and 
\cite{ebrero09} XLFs, where the authors use the luminosity function for the 0.5-2 keV observed band.
The luminosity distribution of the soft band AGNs estimated using A15 is more consistent
with the prediction of AGN XLF by \cite{miyaji2000}. The possible reasons for the disagreement between various AGN XLFs
are described in detail in section-5.3 of \cite{kolodzig13a}.

\subsection{Clustering term}
The clustering term of the AGN angular auto-correlation power spectrum can be further divided into two terms
\citep{mo10, miyaji11, helgason14},
\begin{equation}
  C^{\rm AGN,C}_l=C^{\rm AGN,1h}_{l}+C^{\rm AGN,2h}_{l},
\end{equation}
where $C^{\rm AGN,1h}_{l}$ is due to the correlation between AGNs within the same halo and $C^{\rm AGN,2h}_{l}$ 
is due to the correlation between AGNs residing in different haloes.

\subsubsection{Two-halo clustering term} 
\label{sec-2hc}
The contribution of clustering term to the AGN angular auto-correlation power spectrum due to the correlation between 
AGNs residing in different haloes (under flat sky approximation) is 
\begin{equation}
 C^{\rm AGN,2h}_{l}=\int dz \frac{dV}{dzd\Omega} P^{2h}_{\rm AGN}\Bigl(k=\frac{l}{\chi(z)}\Bigr) [W^{\rm AGN}(z)]^2
 \label{eqn-cl2h}
\end{equation}
where $\chi(z)$ is the comoving distance,
\begin{eqnarray}
 P^{2h}_{\rm AGN}(k) &\approx& P_{\rm lin} (k)  \Bigl[ \int dM\frac{dn}{dM} b_h(M,Z) \times \nonumber\\
 && \Bigl(\frac{\langle N_c(M)\rangle+\langle N_s(M)\rangle f_{\rm AGN}(k,M,z)}{\bar{n}_{AGN}}\Bigr) \Bigl]^2
\label{eqn-p2h}
 \end{eqnarray}
and 
\begin{equation}
 W^{\rm AGN}(z) = \int d\log L_x \phi_{\rm AGN}(L_X) 
 \label{eqn-wagn}
\end{equation}

\begin{figure}
\begin{center}
\includegraphics[width=8.5cm,angle=0.0 ]{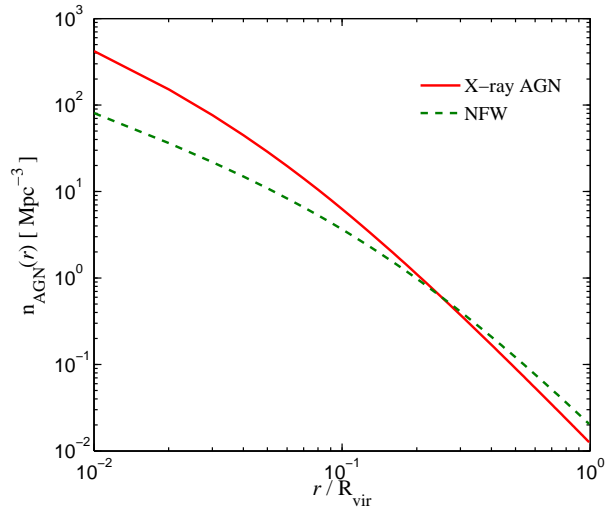}
\end{center}
\caption {The normalized radial distribution of the X-ray AGNs (solid red line) with concentration parameter
given by equation-\ref{eqn-conc}, compared with the NFW profile (dashed green line) 
with concentration parameter from \citet{duffy08},
for a halo mass 
$M_h=10^{14} h^{-1} M_{\odot}$ at $z=0$.}
\label{fig-nrad}
\end{figure}

\begin{table}
\caption{Parameters of MOF.}
\centering 
\begin{tabular}{ c c }
 \hline 
 Parameter & Best fitting value \\[0.5ex]
 \hline \\
 
$\log(M_{\rm min}/h^{-1} M_{\odot})$ & 13.65 \\
$\sigma_{\log M}$ & 0.78 \\
$\log(M_1/h^{-1} M_{\odot})$ & 14.32 \\
$\alpha$ & 2.59 \\
$\log(M_{\rm cut}/h^{-1} M_{\odot})$ & 11.0 \\
\hline

\end{tabular}
\label{tab-mof}
\end{table}

where $P_{\rm lin} (k)$ is the linear matter power spectrum, $b_h(M,z)$ is the linear bias parameter,
$\bar{n}_{\rm AGN}$ is the average comoving number density of the AGNs, 
$\langle N_c(M)\rangle$ and $\langle N_s(M)\rangle$ are the average numbers
of central and satellite 
AGN residing in a halo of mass M, respectively, and $f_{\rm AGN}(k,M,z)$ [defined such that, 
$f_{\rm AGN}(k,M,z)\rightarrow1$ at large scales]
is the Fourier transform of the
normalized AGN distribution within a halo and it is given by,
\begin{eqnarray}
  f_{\rm AGN}(k,M,z)=\int dr \hspace{1mm} 4 \pi r^2 n_{\rm AGN}(r) \frac{\sin(kr)}{kr}
\end{eqnarray}
where $n_{\rm AGN}(r)$ is the radial distribution of the satellite AGN, normalized such that its volume integral within the 
virial radius of the halo is unity.
In figure-\ref{fig-nrad}, we show $n_{\rm AGN}(r)$ for $M_h=10^{14} h^{-1} M_{\odot}$ at $z=0$  
which is assumed to be given by the Navarro-Frenk-White (NFW) profile \citep{nfw97} with a concentration parameter
(R13 and references therein),

\begin{equation}
 c_{\rm AGN}(M,z)=\frac{32}{(1+z)} \Bigl(\frac{M}{M_{\rm ch}}\Bigr)^{-0.13}
 \label{eqn-conc}
\end{equation}
where $M_{\rm ch}$ is a characteristic mass at z=0, defined such that $ \sigma(M_{\rm ch})=1.686$, where 
$\sigma(M)$ is the present day smoothed variance of density fluctuations.
In the same figure, we also show the NFW profile (with concentration parameter from \citealp{duffy08}), 
which describes the radial distribution of the dark matter.
Due to the high concentration of the AGNs, the radial distribution profile
of the AGNs appears steeper than the NFW profile.

Our calculation of the two-halo term does not take into account the effect of scale-dependent bias. 
Also, if the separation between two objects is less than the sum of the virial radii of their host haloes,
they should not be counted in the two-halo term. We neglect this effect, which results in the overestimation 
of the two-halo term at the scale corresponding to the virial radius of the  host halo. However, 
these corrections are expected to be significant only at small scales ($\sim$ virial size of the object), 
where one-halo term becomes more important. Therefore, neglecting this effect is not expected to make much difference 
to our analysis. These corrections are described in detail in \cite{tinker05}, \cite{zheng09}, and
\cite{vanden13} for the real space correlation function.

We use the mean occupation function (MOF) i.e. the number of AGNs ($\langle N(M)\rangle$) residing in a halo of mass M
from R13, where the authors use the measurement of 2PCF
of X-ray selected AGNs by \cite{allevato11} to determine the HOD parameters.
\begin{eqnarray}
 \langle N(M)\rangle &=&\frac{1}{2}\Bigl[1+\rm erf \Bigl (\frac{\log M - \log M_{\rm min}}{\sigma_{\log M}} \Bigr) \Bigr] \nonumber \\
 &&+\Bigl(\frac{M}{M_1}\Bigr)^{\alpha} \exp\Bigl(-\frac{M_{\rm cut}}{M} \Bigr)
 \label{eqn-mof}
\end{eqnarray}

\noindent where erf is the error function, $M_{\rm min}$, $\sigma_{\log M}$, $M_1$, $\alpha$ and $M_{\rm cut}$ are the 
model parameters and their values are shown in table-\ref{tab-mof}.
The first part of equation-\ref{eqn-mof} represents the central AGN contribution, $\langle N_c(M)\rangle$ 
and the second part represents the satellite contribution, $\langle N_s(M)\rangle$.
At low halo masses ($M_h<10^{13.5} h^{-1} M_{\odot}$), the MOF is dominated by the central AGNs, whereas, at
high masses ($M_h>10^{13.5} h^{-1} M_{\odot}$), the satellite component takes over. In the high mass regime,
the number of satellite AGNs is approximately $\propto M^{5/2}$.

\begin{figure}
\begin{center}
\includegraphics[width=8.8cm,angle=0.0 ]{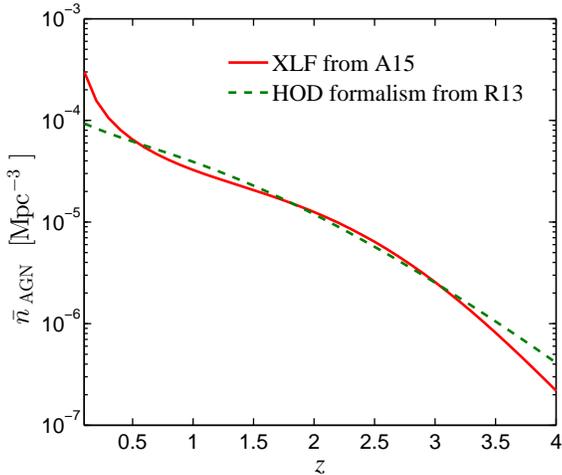}
\end{center}
\caption {Comparison between the X-ray AGN number density in the soft band as a function of redshift estimated
using the XLF from A15 (solid red line)
and the HOD formalism from R13 (dashed green line).}
\label{fig-nagn}
\end{figure}

In figure-\ref{fig-nagn}, we show  the mean number density of AGNs calculated using the 
MOF from R13 ($\bar{n}_{AGN}=\int dM \langle N(M)\rangle \frac{dn}{dM}$) and the
XLF from A15 ($\bar{n}_{AGN}=\int d\log L_X \phi_{\rm AGN}(L_X)$) 
as a function of redshift, 
confirming that there is a good agreement between the two approaches,
especially in the redshift range 0.5-3.
Note that we choose the HOD model by R13 over other models for the X-ray AGNs \citep{miyaji11, allevato12} as the 
X-ray AGN sample used by R13 span a much wider redshift range
compared to that by \cite{miyaji11} ($0.16<z<0.36$) and \cite{allevato12} ($z\lesssim 1$) and is thus 
easier to use with the luminosity function considered in this paper.

\subsubsection{One-halo clustering term} 
The contribution of the clustering term to the AGN angular auto-correlation power spectrum due to the correlation between 
AGNs residing in the same halo can be obtained by replacing $P^{2h}_{\rm AGN}(k)$ by $P^{1h}_{\rm AGN}(k)$ in 
equation-\ref{eqn-cl2h}, with
\begin{equation}
  P^{1h}_{\rm AGN}(k) = \int dM \frac{dn}{dM} \frac{\langle N(M)(N(M)-1)\rangle}{\bar{n}^2_{\rm AGN}} f^2_{\rm AGN}(k)
\label{eqn-p1h}
\end{equation}

Equation-\ref{eqn-p1h} assumes that all AGNs residing in a halo follow the same radial profile \citep{mo98}. 
However, the central AGN is located near the centre of the halo and the satellite AGNs are assumed to follow
the radial distribution shown figure-\ref{fig-nrad}.
Taking into account the different distributions of central and satellite AGNs we get
\begin{equation}
  P^{1h}_{\rm AGN}(k) = P^{1h,ss}_{\rm AGN}(k) + P^{1h,cs}_{\rm AGN}(k)
\label{eqn-p1h-tot}
\end{equation}
where $P^{1h,ss}_{\rm AGN}(k)$ and $P^{1h,cs}_{\rm AGN}(k)$
are the terms due to satellite-satellite and central-satellite correlation{\hspace{-0.2mm}}\footnote{There is a typo
in equation-10 of \cite{miyaji11}. The first term of the integrand 
should be  2 $\langle N_c\, N_s \rangle$.}, respectively.
\begin{equation}
  P^{1h,ss}_{\rm AGN}(k) = \int dM \frac{dn}{dM} \frac{\langle N_s(M)\rangle^2}{\bar{n}^2_{\rm AGN}} f^2_{\rm AGN}(k)
\label{eqn-p1h-ss}
\end{equation}

\begin{equation}
  P^{1h,cs}_{\rm AGN}(k) = \int dM \frac{dn}{dM} \frac{ 2 \langle N_c(M)\rangle \langle N_s(M)\rangle}
  {\bar{n}^2_{\rm AGN}} f_{\rm AGN}(k)
\label{eqn-p1h-cs}
\end{equation}
Here, we have assumed that the satellites follow Poisson distribution
for which $\langle N_s(M)(N_s(M)-1)\rangle=\langle N_s(M)\rangle^2$.

\begin{figure}
\begin{center}
\includegraphics[width=8.5cm,angle=0.0 ]{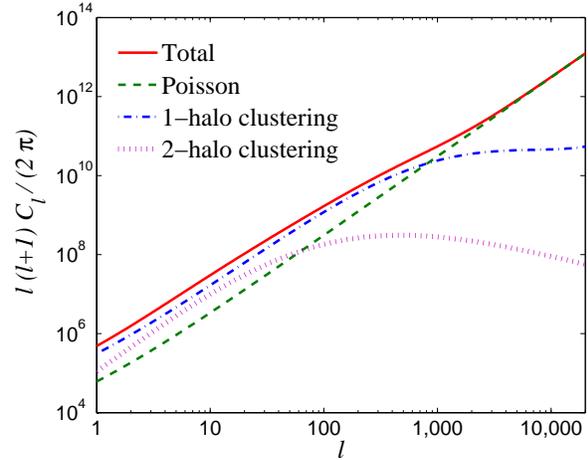}
\end{center}
\caption {The total power spectrum (solid red line), Poisson term (dashed green line), 
one-halo (dot-dashed blue line) and two-halo (dotted magenta line) clustering terms due to the resolved AGNs in the soft band.}
\label{fig-cl}
\end{figure}

In figure-\ref{fig-cl}, we show the number-weighted auto-correlation power spectrum of X-ray AGNs. 
At large $l$-values ($l>1000$)
i.e. small scales, the Poisson term dominates the total power spectrum. However, at small 
$l$-values ($l<1000$) i.e. large angular scales, the clustering is larger than the Poisson term. Within the 
clustering term, the one-halo term is always greater than the two-halo term. Both, the one-halo and the two-halo terms
increase with increasing $l$, till $l\sim 500$, beyond which the two-halo term decreases and the one-halo term continues
to increase till $l\sim 1000$, remains approximately constant till $l\sim 10000$ and dies down with further
increase in $l$.

\subsection{Redshift and mass dependence}
\label{sec-clmz}
\begin{figure*}
\begin{center}
\includegraphics[width=18cm,angle=0.0 ]{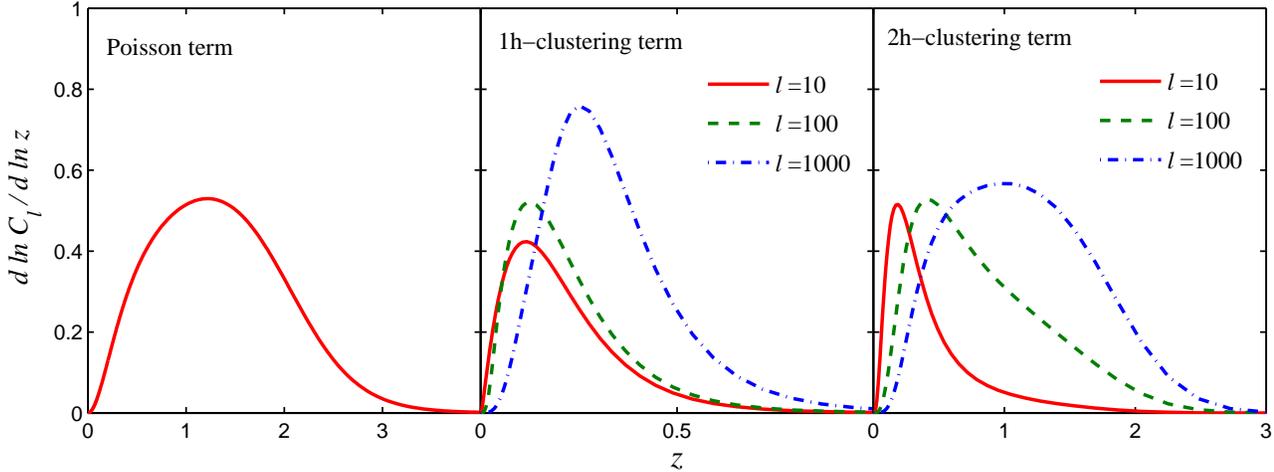}
\end{center}
\caption{The first, second and third
panel show the redshift dependence of the Poisson, one-halo clustering and two-halo clustering terms
of the resolved AGN auto-correlation power spectrum in the soft band, respectively. In the case
of clustering power spectra, $l=$10, 100 and 1000 are shown by solid red, dashed green and dot-dashed blue lines,
respectively.}
\label{fig-clz}
\end{figure*}

We show the redshift dependence of the Poisson, one-halo and two-halo clustering terms
of the AGN auto-correlation power spectrum in figure-\ref{fig-clz}.
Since, the Poisson term is essentially the number of AGNs, its redshift dependence is
identical to the solid red curve in figure-\ref{fig-dndz}, with a peak near $z\sim$ 1.2.
The one-halo clustering term peaks near $z\sim$ 0.1 for $l=$ 10. Increasing the value of $l$ does not significantly change the 
redshift corresponding to the peak. Even at $l=$ 1000, the power spectrum peaks at $z\sim$ 0.25.
Therefore, the one-halo clustering term is dominated by low redshifts where the radial structure of the haloes can be probed.

In the case of two-halo clustering term, the power
spectrum peaks at $z\sim$ 0.2, 0.4 and 1 for $l=$ 10, 100 and 1000, respectively. 
The two-halo term, comes from the correlation of two distinct haloes, and so depends on the
angular difference between two haloes, and hence smaller angular separation (or higher $\ell$-value) peaks at higher z.
For the one-halo, similar to the two-halo clustering
terms, the reason of increasing contribution from high redshift AGN with increasing $l$ value is that the larger $l$ values
correspond to smaller angular scales and the distances at higher redshifts appear smaller on the sky.

\begin{figure}
\begin{center}
\includegraphics[width=12cm,angle=0.0 ]{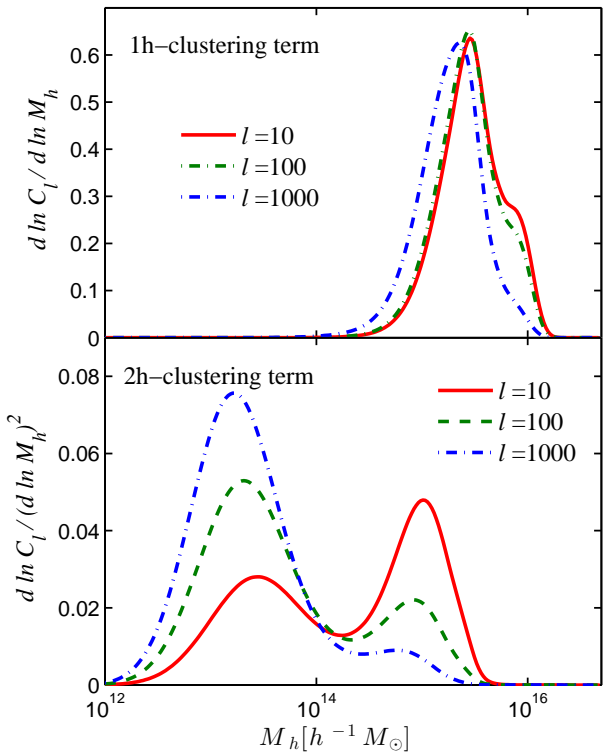}
\end{center}
\caption{The mass dependence of the one-halo (left-hand panel) and two-halo (right-hand panel) clustering
terms of the resolved AGN auto-correlation 
power spectrum in the soft band. The solid red, dashed green and dot-dashed blue lines corresponds to $l=$10, 100 and 1000, 
respectively.}
\label{fig-clm}
\end{figure}

In figure-\ref{fig-clm}, we show the mass dependence of the AGN clustering terms. 
The one-halo term peaks near $2-3 \times 10^{15} h^{-1} M_{\odot}$ in the $l$-range $\sim 10-1000$ and 
the peak shifts only slightly to smaller masses with increasing $l$ as smaller
objects contribute more at large $l$. There is negligible contribution coming 
from low mass haloes ( below $10^{14} h^{-1} M_{\odot}$) to the one-halo term as 
expected from the AGN MOF. Massive  haloes contribute the most due to the large 
number of satellite AGNs residing in them.
However, we note that the mass dependence of the one-halo term is really sensitive
to the relation between the satellite AGN MOF and the host halo mass. Since, the average number of satellite AGNs
increases rapidly with the halo mass ($\langle N_s(M)\rangle \propto M^{2.59}$), 
most of the contribution to the power spectrum comes from the high mass end of the galaxy population
where the number of galactic haloes decreases exponentially.
Also, there is a large uncertainty in the determination of $\alpha$ [hence, in the determination of 
$\langle N_s(M)\rangle$], as shown in R13. 
Therefore, the power spectrum as well as its mass distribution may change
significantly with a different value of $\alpha$.

The two-halo term shows a prominent double peaked structure in its mass
distribution due to the different mass dependences of central and satellite AGNs.
The lower mass peak ($\sim 10^{13} h^{-1} M_{\odot}$) comes from the central 
AGN contribution, whereas, the higher mass peak ($\sim 10^{15} h^{-1} M_{\odot}$) 
comes from the satellite AGN contribution to the two-halo term. The relative strength of 
these two peaks depend on the $l$ value. For small $l$ values (i.e. large scales) major 
contribution comes from the massive haloes, whereas, at large $l$ values (i.e. small scales) 
the low mass haloes rule the two-halo term. Therefore, at small $l$, the higher mass peak is more
prominent than the smaller mass peak and at large $l$ smaller mass peak overshadows the higher mass peak.

\section{Forecast for constraining HOD parameter using auto-correlation power spectrum}
\label{sec-forecast}
There are four free parameters in the AGN HOD model: $M_{\rm min}$, $\sigma_{\log M}$, $\alpha$ and $M_1$,
where $M_{\rm min}$ is the halo mass at which $\langle N_{\rm cen} \rangle = 0.5$, $\sigma_{\log M}$ is the width of
softened step function used for the central AGN MOF, $\alpha$ is slope of power law relation between the satellite
AGN MOF and the halo mass and $M_1$ is the mass scale at which $\langle N_{\rm sat} \rangle \approx 1$.
The AGN auto-correlation power spectrum can be directly related to the HOD parameters through simple power laws 
(varying one parameter at a time and keeping all others fixed)
in a broad range around the fiducial values of these parameters.
For example, $C_l$ has power law dependence on these parameters as follows:
$C_l \propto$ $(\log M_{\rm min})^{1.5}$, $\sigma_{\log M} ^{-2}$ and $(\log M_1)^{-4}$ with little to no change in the power 
law index as a function of $l$. 
However, the variation of $C^{1h} _l$ with $\alpha$ cannot be fit by a single 
power law. The one-halo term has $\langle N _s (M)\rangle ^2$ in its integrand and the satellite MOF itself 
is proportional to $M^{2.59}$. Consequently, the power spectrum is highly biased towards the massive haloes 
as discussed in section-\ref{sec-clmz}. However, varying the value of $\alpha$ can significantly alter the power 
spectrum due to the presence of $\langle N _s (M)\rangle ^2$
term. Increasing the value of $\alpha$ results in a large increase in the power spectrum 
causing the sensitive dependence of the power spectrum on the value of $\alpha$.  
For $0.5< \alpha <1.5$, $C^{1h} _l \propto \alpha^{1.5 (1)}$, whereas 
in the range, $1.5< \alpha <3.0$, $C^{1h} _l \propto \alpha^{8.7 (6.8)}$, at $l=10 (1000)$. 

Similarly, $C^{2h} _l$ is proportional to $(\log M_{\rm min})^{0.5}$ and $\sigma_{\log M} ^{-1}$, 
and the power law indices are nearly independent
of $l$. However, in the case of $\log M_1$, the power law index depends on the fiducial value of 
$M_1$ as well as the value of $l$. At $l=10$, near the fiducial value of $M_1$, $C^{2h} _l \propto(\log M_1)^{-1.1}$.
Analogous to the one-halo term, the relation between the two-halo term and $\alpha$ is really sensitive to
the value of $\alpha$ as well as $l$, with 
 $C^{2h} _l \propto \alpha^{0.5 (0.3)}$ for $0.5< \alpha <2.5$ and $C^{2h} _l \propto \alpha^{2.5 (0.6)}$ 
 for $2.5< \alpha <3$, at $l=10 (1000)$.

Figure-1-(b) of R13 shows $\langle N_c (M)\rangle $ and 
$\langle N _s (M)\rangle $ with their 1-$\sigma$ uncertainties. 
There is a large uncertainty in $\langle N_c (M)\rangle $ at low halo masses ($M_h<10^{12.5} h^{-1} M_{\odot}$)
and in $\langle N _s (M)\rangle $ throughout
the entire mass range. This is due to the current large uncertainties in the determination of 
the model parameters, especially for the satellite AGNs. 
Also, the X-ray AGN sample \citep{allevato11} used to construct the HOD model in R13
spans a broad redshift range (0-4) with a median redshift, $z_{\rm med} \sim 1.2$.
Therefore, this HOD model represents the AGN population at the median redshift, 
an average over the redshift range of the AGN sample and it lacks any redshift dependence.
Other studies (eg. \citealt{koutoulidis13, gatti16}) involving the X-ray AGN clustering measurement also indicate that the HOD 
may have a weak redshift dependence. However, it is not possible to put strong constraints
due to the small sample sizes. In this section, we look into the possibility of constraining
the redshift evolution of the HOD model with upcoming eROSITA all-sky survey using a
Fisher matrix analysis.

For simplicity, we consider a power law redshift dependence of the parameters.
For the purpose of this study, we choose following Fisher parameters:
\begin{equation}
 \Bigl\{ \{ \log(M_{\rm min}), \sigma_{\log M}, \log(M_1), \alpha \} ,\{ \gamma_{M_{\rm min}},
 \gamma_{\sigma_{\log M}}, \gamma_{M_1}, \gamma_{\alpha} \} \Bigr\} \;\; ,
 \label{fischer}
\end{equation}
where the power law indices $\gamma_a$s are defined as,
\begin{equation}
 p_a=p^{\rm fid} _a \Bigl( \frac{1+z}{1+z_{\rm med}} \Bigr)^{\gamma_{a}}
 \label{eqn-fish}
\end{equation}
where, $p^{\rm fid}_a$s are the fiducial values of the HOD parameters (specified in table-\ref{tab-mof})
and $\gamma_{a}$s are the corresponding power law indices. The fiducial values of $\gamma_a$s are
zero i.e. no redshift evolution.

The Fisher matrix can be calculated using,
 \begin{equation}
 F_{ab} \,=\Sigma_{ll'} \,\frac{\partial C_\ell}{\partial p_a} (M_{\ell \ell{^\prime}})^{-1} \frac{\partial C_{\ell{^\prime}}}{\partial p_b} \delta_{ll'}
\end{equation}
where $p_a$s are the model parameters and $M_{\ell \ell{^\prime}}$ is the covariance matrix that
incorporates the uncertainty in the $C_l$s. $M_{\ell \ell{^\prime}}$ is given by,
\begin{equation}
 M_{\ell \ell{^\prime}}=\frac{1}{f_{\rm sky}} \Bigl[ \frac{(C_l + N_l)^2 \delta_{\ell \ell{^\prime}}}{(l+\frac{1}{2}) 
 \Delta l} \Bigr]
\end{equation}
where $f_{\rm sky}$ is the sky coverage of the survey, $N_l$ is the noise in the power spectrum
(we consider AGN shot noise only) and $\Delta l$ is the $l$-bin size.
For simplicity, we neglect the trispectrum contribution to $M_{\ell \ell{^\prime}}$ and take $\Delta l =100$.

We show the forecasted constraints on the parameters in table-\ref{tab-uncertainty}. Note that 
the no priors, fixed HOD and no redshift evolution cases correspond to no priors on any of the Fisher parameters,
no change in the fiducial values of the HOD parameters and no redshift evolution of the HOD parameters, respectively.
Using this method we find that the redshift evolution of the HOD parameters can be constrained to 
$\Delta \gamma_{M_{\rm min}} \sim 0.2$, $\Delta \gamma_{\sigma_{\log M}} \sim 3$, 
$\Delta \gamma_{M_1} \sim 0.02$ and $\Delta \gamma_{\alpha} \sim 0.03$,
with eROSITA and fixed HOD parameters.
Since, the AGN power spectrum is more sensitive to the satellite AGNs compared to the central AGN,
$ \gamma_{M_1} $ and $\gamma_{\alpha}$ are better constrained than
$\gamma_{M_{\rm min}} $, $\gamma_{\sigma_{\log M}}$.
Similarly, in the absence of any redshift evolution, the HOD parameters can be constrained to 
$\Delta\, {\log M_{\rm min}} \sim 1.1$, $\Delta {\sigma_{\log M}} \sim 0.9$, 
$\Delta\, {\log M_1} \sim 0.16$ and $\Delta {\alpha} \sim 0.015$.
However, the uncertainties on these parameters crucially depend on 
the prior information of the model parameters. These constraints degrade significantly if the priors
on other parameters are removed as shown in table-\ref{tab-uncertainty} 
and one cannot obtain any strong constraint in such a situation.

The poor constraints obtained from auto-correlation in the absence of strong priors can be improved by combining 
the information from different probes of the same parameters. Hence, we compute the cross-correlation power
spectrum of X-ray emission from the AGN and the galaxy weak lensing to investigate how this combination can be 
used to improve the constraints on the HOD parameters.

\begin{figure}
\begin{center}
\includegraphics[width=8.5cm,angle=0.0 ]{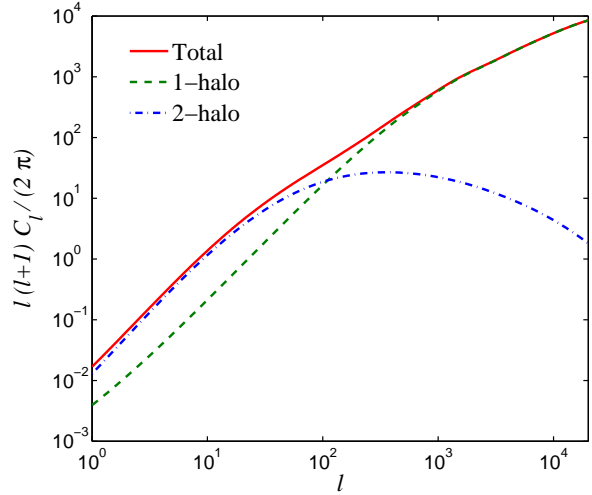}
\end{center}
\caption {The cross-correlation power spectrum of soft X-ray emission from the AGNs and 
weak lensing. The total power spectrum, one-halo and two-halo terms are indicated by
solid red, dashed green and dot-dashed blue lines, respectively.}
\label{fig-cl-lens}
\end{figure}

\begin{figure}
\begin{center}
\includegraphics[width=9.0cm,angle=0.0 ]{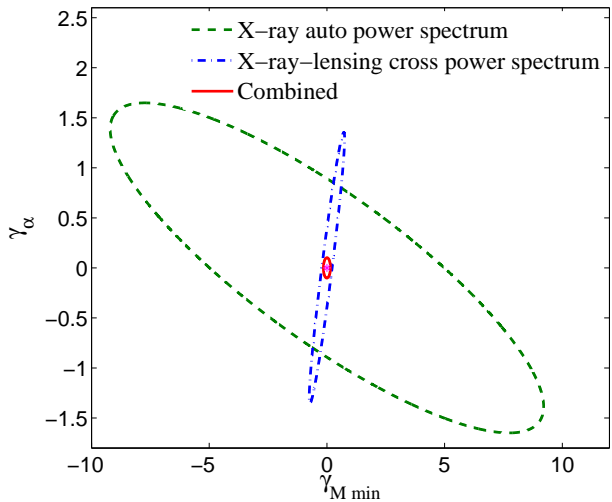}
\end{center}
\caption {The forecasted 68\% confidence limit contours for the redshift evolution parameters of the
HOD model, $\gamma_{M_{\rm min}}$ and 
$\gamma_{\alpha}$. The dashed green and dot-dashed blue contours correspond to the constraints due to the
X-ray AGN auto and X-ray AGN-galaxy lensing cross-correlation power spectra in the soft band,
respectively, whereas, the solid red contour
corresponds to the combined auto and cross-correlation power spectra constraints.}
\label{fig-fisher}
\end{figure}

\begin{table*}
\caption{Forecasted constraints on the HOD parameters with eROSITA and eROSITA-LSST combination.}
\centering
\begin{tabular}{c c c c c c c c c c c}
\hline \\
  & Power spectrum & $\Delta \log(M_{\rm min})$ & $\Delta \sigma_{\log M}$  & $\Delta \log(M_1)$  & $\Delta \alpha$ 
  & $\Delta \gamma_{M_{\rm min}}$ & $\Delta \gamma_{\sigma_{\log M}}$ & $\Delta \gamma_{M_1}$ & $\Delta \gamma_{\alpha}$\\\\
 \hline
 & X-ray auto & 35 & 23 & 6 & 3 & 6 & 75 & 1 & 1\\[-0.5ex]
 No priors \Bigg \{& X-ray-lensing cross  & 2.5 & 1.5 & 1 & 1.2 & 0.5 & 5.7 & 0.16 & 0.9\\[-0.5ex]
 & Auto+Cross & 0.6 & 0.3 & 0.17 & 0.13 & 0.1 & 0.9 & 0.02 & 0.07\\[-0.5ex]\\\\

  & X-ray auto & - & - & - & - & 0.2 & 3 & 0.02 & 0.03\\[-0.5ex]
 Fixed HOD \Bigg \{&X-ray-lensing cross & - & - & - & -& 0.1 & 1 & 0.03 & 0.11 \\[-0.5ex] 
 & Auto+Cross & - & - & - & - & 0.02 & 0.2 & 0.004 & 0.008 \\[-0.5ex]\\\\
 
 & X-ray auto & 1.1 & 0.9 & 0.16 & 0.015 & - & - & - & -\\[-0.5ex]
 No redshift evolution \Bigg \{ &X-ray-lensing cross & 0.4  & 0.2 & 0.13 & 0.13 & - & - &- &-\\[-0.5ex]
 & Auto+Cross & 0.1 & 0.06 & 0.02 & 0.01 & - & - &- &-\\[-0.5ex]
 \hline
 \end{tabular}
 \label{tab-uncertainty}
\end{table*}

\section{Cross-correlating X-ray AGN with lensing}
\label{sec-kxcross}
When the light from a background galaxy is bent by a structure along the line-of-sight,
the resulting distortion is known as gravitational lensing. There are two effects of lensing
on the image of a background galaxy: convergence ($\kappa$), which represents the isotropic stretching of the image, and
shear ($\gamma$), which represents the anisotropic stretching thus distorting a circular image into an elliptical one.
In the weak lensing regime both the convergence and shear are small i.e. $\kappa \ll 1$, $\gamma \ll 1$. The convergence
field is a direct probe of the gravitational potential of the lens mass and it can be obtained from the measurement of the 
shear field. For the detailed procedure of measurement of the shear field and construction
of convergence maps see e.g. \cite{refregier03}, \cite{kilbinger15} and references therein.

Cross-correlation of galaxy lensing with other probes such as Sunyaev-Zel'dovich effect \citep{batt15, ma14},
 cosmic microwave background lensing \citep{hand14}, has been used to understand baryonic physics as well as to put 
 constraints on the cosmological parameters.
Since, the motivation of the HOD formalism is to establish a connection between
the AGNs and the host halo, cross-correlating X-ray emission from the AGNs and the 
lensing signal can provide an additional tool to constrain the HOD model.
Therefore, in this section, we estimate the cross-correlation power spectrum of the lensing convergence field with
the X-ray emission from AGNs.

\subsection{X-ray-lensing cross-correlation power spectrum}
Analogous to the X-ray auto-correlation power spectrum, the AGN X-ray-lensing cross-correlation power spectrum 
in the thin lens limit (i.e. the thickness of the lens is much smaller than 
the distance between the lens and observer as well as the distance between the lens and source) is given by,
\begin{equation}
 C^{\rm AGN,\kappa}_l = \int dz \frac{dV}{dzd\Omega} W_{\rm AGN}(z) W_{\kappa}(z) P_{\rm AGN,\kappa}\Bigl(k=\frac{l}{\chi(z)}\Bigr)
\end{equation}
\\ where $W_{\rm AGN}(z)$ is given by equation-\ref{eqn-wagn}. The lensing kernel $W_{\kappa}(z)$ 
for a flat Universe \citep{van14} is given by,
\begin{equation}
 W_{\kappa}(z)=\frac{3}{2}\Omega^0_M \Bigl(\frac{H_0}{c} \Bigr)^2 g(z) (1+z)/\chi (z)
\end{equation}
Here, $g(z)$ is defined as, 
\begin{equation}
  g(z)=\int_\chi^{\chi_H} d\chi' p_s(\chi')\frac{\chi'-\chi}{\chi'}=\int_z ^{z_H} dz' p_s(z') \frac{\chi'-\chi}{\chi'}
\end{equation}
where $p_s(z)=n(z)/n(z=0)$ is the normalized source redshift distribution function. 
Again, the power spectrum can be decomposed into one-halo and two-halo terms.
\begin{equation}
 P_{\rm AGN,\kappa}(k)=P^{1h}_{\rm AGN,\kappa}(k)+P^{2h}_{\rm AGN,\kappa}(k)
\end{equation}
where,
\begin{eqnarray}
 P^{1h}_{\rm AGN,\kappa}(k) &=& \int dM \frac{dn}{dM} f_{\kappa}(k,M,z)  \times \nonumber\\
 &&  \Bigl[ \frac{\langle N_c(M)\rangle+\langle N_s(M)\rangle f_{\rm AGN}(k,M,z)}{\bar{n}_{\rm AGN}} \Bigr]
 \label{eqn-clak1h}
\end{eqnarray}
and
\begin{eqnarray}
 P^{2h}_{\rm AGN,\kappa}(k) \approx P_{\rm lin}(k) \Bigl[ \int dM \frac{dn}{dM} b_h(M,Z) f_{\kappa}(k,M,z) \Bigr]  \nonumber\\
  \times \Bigl[ \int dM \frac{dn}{dM} b_h(M,Z)
  \frac{\langle N_c(M)\rangle+\langle N_s(M)\rangle f_{\rm AGN}(k,M,z)}{\bar{n}_{\rm AGN}} \Bigr]
\end{eqnarray}
 
where, $f_{\kappa}(k,M,z)$ is the Fourier transform of the dark matter density profile and its given by,
\begin{equation}
 f_{\kappa}(k,M,z)=\int 4 \pi r^2 dr \frac{\rho_{\rm DM}(r)}{\bar{\rho}_M} \frac{\sin(kr)}{kr}\\
\end{equation}

where $\rho_{\rm DM}(r)$ is the NFW density profile and
$\bar{\rho}_M$ is the comoving matter density of the universe.

We choose the combination of eROSITA and LSST surveys to forecast the constraints on the HOD model
that can be obtained from such a study. The reason of this choice is the high sky coverage of the 
overlapping region of the two surveys.
LSST is a ground based optical telescope, presently under construction. 
We take the sky coverage of the overlapping region between eROSITA and LSST, $f_{\rm sky} \sim 0.5$, for the 
calculation of the covariance matrix.
In order to calculate the lensing part, we take the source redshift distribution function,
$p_s(z)=(z^3/2z^3 _0) e^{-z/z_0}$ \citep{batt15}, where $z_0 =1/3$ for the LSST survey. 

In figure-\ref{fig-cl-lens}, we show the one-halo, two-halo and total X-ray AGN-lensing 
cross-correlation power spectrum. 
There is an interesting difference between the X-ray auto-correlation power spectrum and the
X-ray-lensing cross-correlation power spectrum.
In the case of X-ray auto-correlation power spectrum, two-halo term is always smaller than the one-halo term.
But for the cross-correlation power spectrum, on large scales ($l \lesssim 100$),
the two-halo term is larger than the one-halo term.
This behavoir is due to the reason that in the case of galaxy weak lensing auto-correlation
power spectrum, the one-halo term becomes larger than the two-halo term 
at comparatively larger multipoles ($l>$ 100, see for example \citealt{takada07}).
As a result, the X-ray-lensing cross-correlation power spectrum shows the transition from two-halo to one-halo dominance
at intermediate multipoles.

\subsection{Forecast using the cross-correlation power spectrum}
The uncertainty in the X-ray-lensing cross-correlation power spectrum is given by, 

\begin{equation}
 M^{\rm AGN, \kappa}_{ll'} = \frac{\delta_{ll'}}{f_{\rm sky}(2l+1)\Delta l} \times \Bigl[\hat{C}^{\rm AGN, AGN}_{l}
 \hat{C}^{\kappa \kappa}_{l}+\hat{C}^{\rm AGN, \kappa}_{l} \hat{C}^{\rm AGN, \kappa}_{l}\Bigr]
 \label{eqn-mll}
\end{equation}

where, $\hat{C}^{ij}_{l}$s are the auto ($i=j$) or cross ($i\neq j$) power spectra including noise.
We consider only the shot noise term to calculate the noise in the 
X-ray auto-correlation power spectrum, as mentioned in section-\ref{sec-forecast}. 
Noise in the lensing auto-correlation power spectrum is, $N^{\kappa \kappa} _l = \sigma^2 _{\gamma}/n_s$ \citep{batt15},
where, $\sigma_{\gamma}$ and $n_s$ are the intrinsic ellipticity dispersion per component and
the 2-dimensional angular number density of the source
galaxies, respectively. For LSST, $\sigma^2 _{\gamma}=0.28$ and $n_s=40\,\, \rm arcmin^{-2}$.
Assuming that the noise in X-ray and lensing surveys are independent of each other, $N^{\rm AGN, \kappa} _l=0$.

\begin{figure}
\begin{center}
\includegraphics[width=9.5cm,angle=0.0 ]{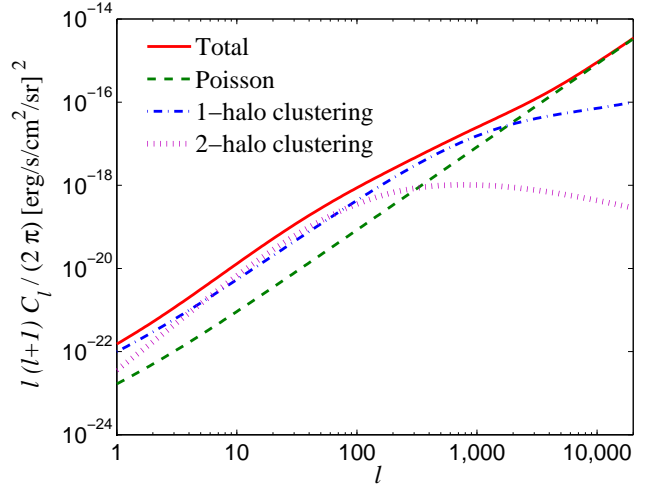}
\end{center}
\caption {Same as figure-\ref{fig-cl} but for the unresolved AGNs.}
\label{fig-cl-unresolved}
\end{figure}

\begin{figure*}
\begin{center}
\includegraphics[width=18cm,angle=0.0 ]{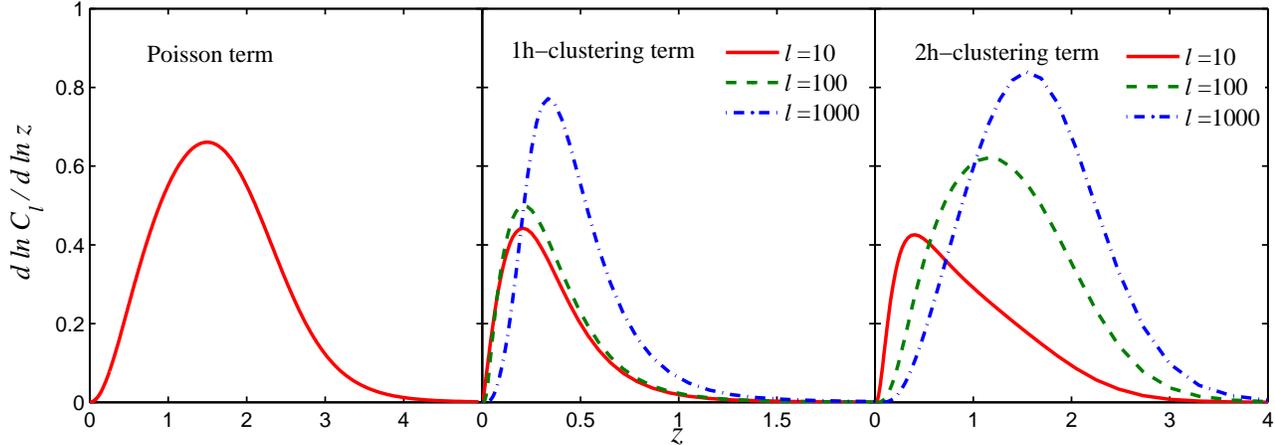}
\end{center}
\caption{Same as figure-\ref{fig-clz} but for the unresolved AGNs.}
\label{fig-clz-un}
\end{figure*}

We choose $\gamma_{M_{\rm min}}$ and $\gamma_{\alpha}$ that represent the central and satellite AGNs, respectively, 
and show the  individual as well as the joint constraints on $\gamma_{M_{\rm min}}$ and $\gamma_{\alpha}$ 
coming from the auto and cross-correlation power spectra in figure-\ref{fig-fisher}.
Note that, to examine how well these spectra combinations can constrain the model parameters,
we do not put any priors on any of the Fisher parameters. 
The X-ray-lensing cross-correlation analysis can alone constrain $\gamma_{M_{\rm min}}$ and $\gamma_{\alpha}$ 
to approximately, 0.5 and 0.9, respectively. The uncertainty on $\gamma_{M_{\rm min}}$ is much better for the cross
correlation signal compared to the auto-correlation signal, whereas, the uncertainty on $\gamma_{\alpha}$
is similar for the two cases.
The reason of the loose constraint on $\gamma_{M_{\rm min}}$ as compared to $\gamma_{\alpha}$ 
coming from the X-ray auto-correlation  power spectrum
 is that the X-ray auto-correlation power spectrum is much more sensitive 
to the satellite AGN MOF, hence $\gamma_{\alpha}$ in comparison to the central AGN MOF, hence $\gamma_{M_{\rm min}}$
(see equation-\ref{eqn-p1h-ss} and \ref{eqn-p1h-cs}).
On the contrary, the X-ray-lensing cross-correlation power spectrum has similar dependence on the central as well as
satellite AGN MOF (see equation-\ref{eqn-clak1h}),
resulting in similar constraints on the corresponding Fisher parameters.
In addition, this difference in dependence of the auto and cross-correlation power spectra on the central and satellite terms
give rise to distinct directions of degeneracy in $\gamma_{M_{\rm min}}$ and $\gamma_{\alpha}$ 
uncertainty contours, as shown in figure-\ref{fig-fisher}. Consequently, the constraints obtained by 
combining the X-ray auto and X-ray-lensing cross-correlation power spectra Fisher matrices are much better than their
individual constraints. 
The constraints obtained from the combined as well as individual Fisher matrices are shown in
table-\ref{tab-uncertainty}.

\section{Unresolved AGNs}
\label{sec-unresolved}
Depending on the flux limit of the X-ray survey, a fraction of AGNs may remain unresolved
in the X-ray map. These unresolved X-ray AGNs are the prime source of contamination while studying 
the diffuse X-ray emission from the hot gas in the ICM and CGM.
Therefore, modelling the unresolved AGNs appropriately is essential to extract the X-ray emission from 
the hot gas and hence the gas physics.
In this section, we estimate the flux-weighted angular power spectrum, the redshift and 
halo mass dependence of the power spectrum
for the unresolved AGNs, assuming that these AGNs are also described by the same luminosity function and the HOD model. 
The reason for the choice of flux-weighted power spectrum is that it is not possible
to identify the source of these X-ray photons as their hosts are not resolved by the X-ray telescope,
which rules out the number-weighted power spectrum. 

\begin{figure}
\begin{center}
\includegraphics[width=12cm,angle=0.0 ]{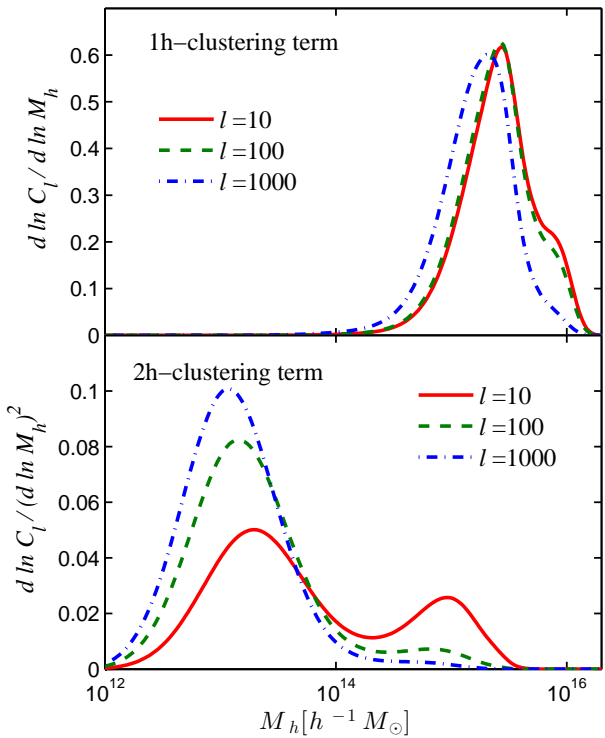}
\end{center}
\caption{Same as figure-\ref{fig-clm} but for the unresolved AGNs.}
\label{fig-clm-un}
\end{figure}

\subsection{Flux-weighted angular auto-correlation power spectrum}
\label{sec-autops-un}
As mentioned in section-\ref{sec-resolved}, the estimator of unresolved AGN power spectrum is given
by equation-\ref{eqn-estim}, where, the signal $x(\theta,\phi)$ is the unresolved X-ray flux.
Analogous to the resolved AGNs, the flux-weighted power spectrum of the unresolved AGNs is the sum of Poisson and 
clustering terms. The Poisson term is given by,

\begin{equation}
 C^{\rm AGN}_{P}=\int dz \frac{dV}{dzd\Omega} \int d\log L_X \phi_{\rm AGN}(L_X,z) S^2 _X
 \label{eqn-clp-un}
\end{equation}
In contrast to the resolved AGN, the lower luminosity limit in the above integral is set to zero and the upper 
luminosity limit is fixed by the sensitivity limit of the X-ray telescope for the unresolved AGN.

Again, the clustering term is composed of one-halo and two-halo terms given by equation-\ref{eqn-p1h} and \ref{eqn-p2h}, 
respectively, with $W^{\rm AGN}(z)$ replaced by,
\begin{equation}
 W^{\rm AGN}(z) = \int d\log L_x \phi_{\rm AGN}(L_X) S_X 
 \label{eqn-wagn-un}
\end{equation}

In figure-\ref{fig-cl-unresolved}, we show the flux-weighted auto-correlation power spectrum for AGNs
lying below the flux resolution
limit of eROSITA. Similar to the resolved AGNs, at lower multipoles, the clustering term is 
dominant whereas at large multipoles, the Poisson term takes over the total power spectrum. However, 
this takes place near $l\sim$ 2000 in the case of unresolved AGNs, compared to $l\sim$ 1000 for the resolved AGNs.
Another visible difference between the resolved and the unresolved AGN power spectrum is that the two-halo
clustering term is always smaller than the one-halo clustering term for the resolved AGN, whereas, the two-halo term is slightly
larger than one-halo term in the $l$-range 10-100 for the unresolved AGNs.

\subsection{Redshift and mass dependence}
In figure-\ref{fig-clz-un}, we show the redshift dependence of the Poisson, one-halo 
clustering and two-halo clustering power spectra
of the unresolved AGNs. 
The redshift distribution of the unresolved AGN power spectra 
have small but non-negligible contribution from the higher redshift AGNs compared to the resolved
AGNs, with similar shape of the overall distribution.
The Poisson term peaks at $z\sim$ 1.5. This peak is at slightly
higher redshift compared to the redshift
distribution of the resolved AGNs (see figure-\ref{fig-dndz}) which peaks near $z \sim$ 1.2,
due to the simple fact that a high redshift AGN with the same luminosity as a low redshift
AGNs, corresponds to a lower flux and hence may remain unresolved depending on the telescope's resolution.
For the one-halo clustering
term, the power spectrum peaks near $z \sim$ 0.2, for $l \sim$ 10 and the peak shifts only slightly to higher redshifts with
increasing value of $l$. 

The two-halo clustering term peaks at $z\sim$ 0.5, for $l \sim$ 10 and the peak shifts significantly 
with increasing value of $l$. For $l \sim$ 1000, the two-halo clustering power spectrum is dominated by $z\sim$ 1.5
AGNs. Since, the total power spectrum is controlled by clustering term at $l<2000$ and Poisson term at $l>2000$, the 
overall power spectrum is governed by low to intermediate redshift AGNs ($z\sim$ 0.2-1.5).

The mass dependence of the unresolved AGN power spectrum is shown in
figure-\ref{fig-clm-un}. The one-halo clustering term of the unresolved and resolved AGNs 
(figure-\ref{fig-clm}) have almost identical halo mass
dependence, with a peak near $10^{15} h^{-1} M_{\odot}$ due to the highly biased MOF of 
the satellite AGNs towards massive haloes. 

The two-halo clustering power spectrum 
also show a double peaked mass distribution similar to the one observed in case of resolved AGNs.
However, the low mass peak 
($\sim 10^{13} h^{-1} M_{\odot}$) is more prominent than the high mass peak ($\sim 10^{15} h^{-1} M_{\odot}$), 
even at $l=10$. At higher $l$ values, most of the contribution to the two-halo term comes from 
$10^{12} - 10^{14} h^{-1} M_{\odot}$ haloes.

\section{Conclusions}
\label{sec-conc}
In this work, we divided the AGNs into two categories: 1) resolved AGNs, which lie above the flux resolution
limit of eROSITA, and 2) unresolved AGNs, lying below this limit. We computed the number-weighted and flux-weighted
 angular power spectra for the resolved and unresolved AGNs, respectively,
in the soft X-ray band. We used
the LADE model for the X-ray AGN luminosity function described in
A15 (which matches well with the observed AGN number counts)
and the HOD model for the X-ray AGNs 
by R13, which describe the luminosity distribution and halo mass dependence 
of the AGNs, respectively. We also calculated the luminosity and redshift dependence of the resolved
AGNs finding that the maximum contribution to the AGN luminosity and redshift distribution comes from
$L_X \sim 2-3 \times 10^{44} \rm \, erg\,\, s^{-1}$ and $z\sim 1.2$, respectively.
We computed the number density of the X-ray AGNs using the XLF as well as the
HOD formalism and found that the two approaches are in good agreement with each other in a wide range 
of redshift (especially at $0.5 \lesssim z \lesssim 3.5$). The X-ray auto-correlation power spectrum has the following features.

\begin{enumerate}

\item The power spectrum of the resolved AGNs is dominated by low redshift AGNs ($z \sim 0.1-0.2$) at low multipoles 
 ($l<1000$) due to the dominant
 contribution from the one-halo clustering term at these multipoles.
 However, at large multipoles ($l>1000$), where the Poisson term controlls 
 the total power spectrum, the redshift dependence of the power spectrum peaks at intermediate redshifts ($z \sim 1.2$).
 The one-halo term is larger than the two-halo term down to $l=1$, though at low multipoles
 these two terms have similar amplitudes.
 
 \item For the unresolved AGNs, the power spectrum is dominated by Poisson term and hence $z \sim 1.5$ AGNs at $l>2000$.
 In the range $l \sim 100-2000$, where the one-halo clustering term is dominant, the total power spectrum is 
 governed by $z \sim 0.2-0.3$ AGNs.
 At $l<100$, the power spectrum has major contribution from $z<1$ AGNs due to significant
 contribution from two-halo as well as the one-halo clustering terms. 
 
 \item The HOD model that we use in this work predicts a large number of satellite AGNs in massive haloes. This mass
 dependence shows up in the mass dependence of the AGN power spectrum (resolved as well as unresolved). 
 Both the one-halo and two-halo 
 terms show peaks at the high mass end ($M_h \sim 10^{15} h^{-1} M_{\odot}$) of the DMH. Interestingly, the 
 two-halo term shows an additional peak at lower halo mass ($M_h \sim 10^{13} h^{-1} M_{\odot}$), 
 which is due to the contribution from 
 the central AGN. The low mass peak becomes more prominent than the high mass peak
 at large multipoles i.e. small angular scales, especially
 in the case of unresolved AGNs.
\end{enumerate}

We also investigated the role of eROSITA in constraining the redshift evolution 
of the HOD parameters, which play a crucial role in determining the AGN
power spectrum, using the Fisher matrix analysis.
We assumed a simple power law dependence of the HOD parameters on the redshift and 
found that the uncertainties in the determination of these power law indices 
vary significantly with and without any priors on other HOD parameters. Without any priors, the X-ray auto-correlation power 
spectrum poorly constraints these parameters (shown in table-\ref{tab-uncertainty}).
To improve upon this, we included X-ray-lensing cross-correlation power 
spectrum, which is motivated by the fact that lensing traces the total halo mass and that the HOD model 
describes the relation between the host halo mass and AGNs.
The cross-correlation power spectrum can alone put much better constraints on the model parameters
(except for the satellite AGN parameters in some situations)
and combining the auto and cross-correlation power spectra improves the uncertainties further.
For example, $\gamma_{M_1}$, which describes the redshift evolution of the relation between the 
satellite AGN MOF and the host halo mass, can be constrained to $\Delta \gamma_{M_1} \sim 1$ and 0.16, using
the X-ray auto and X-ray-lensing cross-correlation power spectra, respectively, and the constraint improves to 
$\Delta \gamma_{M_1} \sim 0.02$ with the auto and cross-correlation power spectra combination, without
any prior information on any other parameter. Adding prior information on the Fisher parameters can significantly
reduce their uncertainties. The uncertainty on $\gamma_{M_1}$ coming from the X-ray auto-correlation power spectrum only,
reduces to $\Delta \gamma_{M_1} \sim 0.03$, if the uncertainties on the fiducial values of the HOD model
parameters are reduced to zero. Other parameters also follow the same trend depending on the sensitivity of the 
power spectrum considered, with respect to the parameter.

Our power spectrum analysis suggest that the present constraints on the AGN HOD parameters will 
improve significantly with the availability of eROSITA and LSST. At the same time, due to the potential of eROSITA
to probe the hot gas in galaxies, such an analysis is expected to play a crucial role
in separating the contribution from the AGNs and the hot gas.\vspace{10mm} \\\\
{\bf{ACKNOWLEDGEMENTS}}\\
We thank the anonymous referee for valuable suggestions and comments.
We thank Takamitsu Miyaji, James Aird, Aseem Paranjape, Suchetana Chatterjee and Jonathan Richardson 
for helpful discussions. We also thank James Aird for providing the results on the uncertainties in the LADE model
parameters.

\footnotesize{
\bibliography{bibtexcgm}{}
\bibliographystyle{mn2e}
}

\end{document}